\documentclass[preprint]{aastex631}

\usepackage{comment}
\usepackage{graphicx}
\usepackage{dcolumn}
\usepackage{bm}
\usepackage{color}

\usepackage{rotating}

\usepackage{threeparttable}

\usepackage{here}

\begin{document}

\title{Low-$J$ transitions in $\tilde{A}^2\Pi(0,0,0)-\tilde{X}^2\Sigma^+(0,0,0)$ band of buffer-gas-cooled CaOH}

\author[0000-0003-0027-0556]{Yuiki Takahashi}
\affiliation{Division of Physics, Mathematics, and Astronomy, California Institute of Technology, Pasadena, California 91125, USA}
\author{Masaaki Baba}
\affiliation{Department of Chemistry, Graduate School of Science, Kyoto University, Kyoto 606-8502, Japan}
\affiliation{Molecular Photoscience Research Center, Kobe University, Kobe 657-8501, Japan}
\author{Katsunari Enomoto}
\affiliation{Department of Physics, University of Toyama, Toyama, 930-8555, Japan}
\author{Ayami Hiramoto}
\affiliation{Research Institute for Interdisciplinary Science, Okayama University, Okayama, 700-8530, Japan}
\author{Kana Iwakuni}
\affiliation{Institute for Laser Science, University of Electro-Communications, 1-5-1 Chofugaoka, Chofu, Tokyo 182-8585, Japan}
\author[0000-0003-3120-2572]{Susumu Kuma}
\affiliation{Atomic, Molecular and Optical Physics Laboratory, RIKEN, 2-1 Hirosawa, Wako, 351-0198, Saitama, Japan}
\author{Reo Tobaru}
\affiliation{Research Institute for Interdisciplinary Science, Okayama University, Okayama, 700-8530, Japan}
\author[0000-0002-0567-6881]{Yuki Miyamoto}
\email{miyamo-y@cc.okayama-u.ac.jp}
\affiliation{Research Institute for Interdisciplinary Science, Okayama University, Okayama, 700-8530, Japan}

\begin{abstract}

Calcium monohydroxide radical (CaOH) is receiving an increasing amount of attention from the astrophysics community as it is expected to be present in the atmospheres of hot rocky super-Earth exoplanets as well as interstellar and circumstellar environments. Here, we report the high-resolution laboratory absorption spectroscopy on low-$J$ transitions in $\tilde{A}^2\Pi(0,0,0)-\tilde{X}^2\Sigma^+(0,0,0)$ band of buffer-gas-cooled CaOH. In total, 40 transitions out of the low-$J$ states were assigned, including 27 transitions which have not been reported in previous literature. The determined rotational constants for both ground and excited states are in excellent agreement with previous literature, and the measurement uncertainty for the absolute transition frequencies was improved by more than a factor of three. This will aid future interstellar, circumstellar, and atmospheric identifications of CaOH. The buffer-gas-cooling method employed here is a particularly powerful method to probe low-$J$ transitions and is easily applicable to other astrophysical molecules.

\end{abstract}

\section{Introduction} \label{sec:intro}

Calcium monohydroxide molecules ($^{40}$Ca$^{16}$O$^{1}$H) now have a special standing because of their expected presence in the atmospheres of hot rocky super-Earth exoplanets (\cite{Bernath2009,Rajpurohit2013, Tennyson2017}). Determining accurate spectroscopic parameters and transition frequencies on relatively simple molecules present in this environment are necessary for exploration of hot rocky super-Earths. Because of highly abundant Ca atoms, CaOH could also be found in interstellar and circumstellar spaces, where the temperature is sometimes in the 10s of kelvin regime. In fact, CaOH has been detected in cool stellar atmospheres at low spectral resolution (\cite{Rajpurohit2016}). CaOH is therefore an excellent spectroscopic target species.

The recent Measured Active Rotational-Vibrational Energy Levels (MARVEL) algorithm (\cite{CSASZAR2007, Furtenbacher2007, Furtenbacher2012, Tobias2019}) has provided a data set of rovibronic (rotation-vibration-electronic) transition frequencies and energy levels of CaOH (\cite{Wang2020}). It evaluated all available spectroscopic data on CaOH from the published literature at the time (\cite{Hilborn1983, Bernath1984, Bernath1985, Coxon1991, Coxon1992, Coxon1994, Jarman1992, li_laser_1992, Li1995, Ziurys1992, Ziurys1996, Scurlock1993, Dick2006}). The uncertainty of final MARVEL data is typically $\geq$ 0.005 cm$^{-1}$ (150 MHz), and as there is an absence of experimental transition frequency values in literature, $J$ = 1/2 energy levels in the $\tilde{A}^2\Pi$ ground vibrational state is missed\footnote{Note that the transitions out of this state have been previously observed for Stark splitting measurements (\cite{steimle_supersonic_1992})}.  Low-$J$ transitions are generally more difficult to characterize with high accuracy at room temperature due to the lower population and, therefore, lower sensitivity. Hence, recent high-level ab initio calculations used $J$ = 3/2 states data for the adjustment of parameters of potential energy surfaces (\cite{Owens2021}).

Several experimental studies have measured the rovibronic spectrum of $\tilde{A}^2\Pi(0,0,0)-\tilde{X}^2\Sigma^+(0,0,0)$ band of CaOH. 
Previous experiments (\cite{Hilborn1983, Bernath1985, steimle_supersonic_1992, Dick2006}) conducted high-resolution spectroscopy using either heated gas or supersonic jet. \cite{kozyryev_determination_2019-2} has reported the transition frequency from $J$ = 3/2 state employing the buffer-gas-cooling (BGC) (\cite{Maxwell2005,Hutzler2012,Yuiki2021}) method with an uncertainty of $\sim$0.03 cm$^{-1}$.

The cryogenic BGC is a powerful method to produce cold molecules in large quantity, regardless of the reactivity and complexity of target species. Using this method, a variety of cold molecular species have been produced and explored in a range of spectroscopy (\cite{Messer1984,Spaun2016,Porterfield2019, Santamaria2016}) as well as precision measurement stuidies for fundametal physics (\cite{ACME2018}) and laser cooling of molecules (\cite{Tarbutt2019,McCarron2018}).  At the Doyle group at Harvard, cold and ultracold experiments have been conducted on CaOH utilizing low-J transitions for laser slowing, cooling, and trapping (\cite{kozyryev_determination_2019-2, baum_1d_2020-1, baum_establishing_2020}). In fact, the 3D Magneto-Optical trapping and subsequent sub-Doppler cooling of buffer-gas-cooled CaOH has been recently achieved using BGC as a pre-cooling technique (\cite{Vilas2022}). From the spectroscopic point of view, the BGC technique is especially suitable for probing low-$J$ transitions because the molecular population is moved into lower-$J$ levels, drastically enhancing the sensitivity. Since the BGC technique relies on mechanism different from the standard methods, such as supersonic jet, to cool the samples, it can serve as a complementary technique to explore species that are challenging to probe using the standard methods, e.g, larger and more complex molecules. (\cite{Spaun2016, Changala2019, Miyamoto2022arxiv}) 

In this study, we explore high-resolution laboratory absorption spectroscopy of the $\tilde{A}^2\Pi(0,0,0)-\tilde{X}^2\Sigma^+(0,0,0)$ transition of CaOH in a cold buffer-gas cell. The BGC method enables us to probe low-$J$ transitions with small Doppler broadening and high sensitivity.  A combination of BGC and co-recorded Doppler-free I$_2$ spectra allows the accurate determination of transition frequencies with an uncertainty of \textless 30 MHz, an improvement of more than three times over previous studies.

\section{Experimental apparatus} \label{sec:apparatus}

\begin{figure}[ht]
\begin{center}
\includegraphics[width=0.48\textwidth]{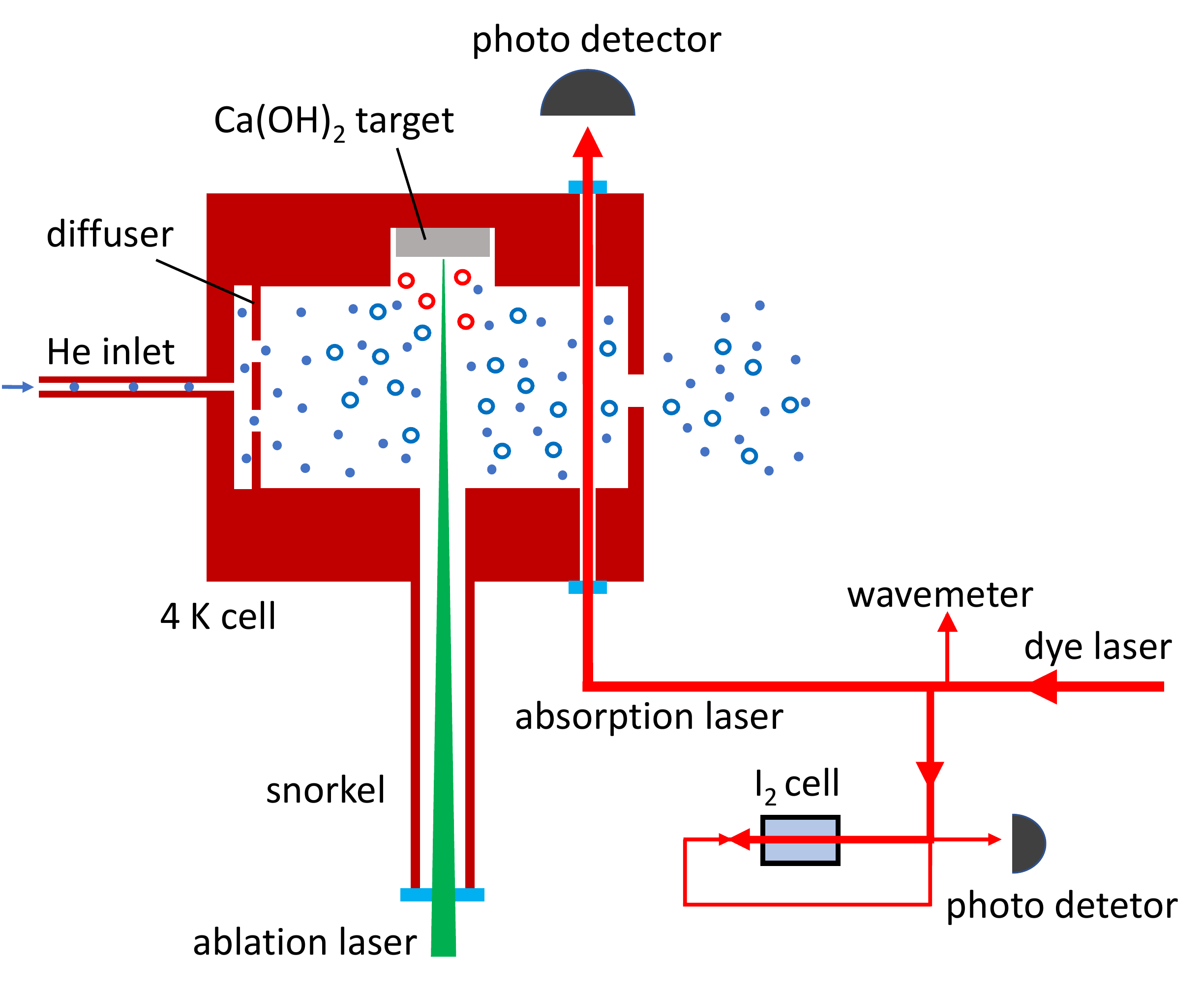}
\end{center}
\caption{Experimental setup for high-resolution spectroscopy of buffer-gas-cooled CaOH. The buffer gas cell bore is cylindrically symmetric. The blue circle shows the helium buffer gas atom whereas red (blue) rings indicate hot (cold) CaOH molecules. The red line indicates the laser path emitted from the ring dye laser. The thickness of the lines indicate the laser power although it is not scaled. Note that the structure of the diffuser is simplified for easier readability.}
\label{cell}
\end{figure}

The cryogenic buffer-gas source for high-resolution spectroscopy on CaOH is shown in Fig. \ref{cell}. The pulsed Nd:YAG laser (532 nm wavelength, $\sim$10 ns wide, $\sim$20 mJ energy, and 10 Hz repetition rate) is focused on and ablates a solid Ca(OH)$_2$ powder target inside a copper cell to produce cold CaOH molecules. The diameter of the focused ablation laser beam at the target is $\sim$100 $\mu$m. The body of the buffer-gas cell is made of a copper block with a cylindrical cavity (5 cm long and 2.5 cm diameter). Ablation laser beam is sent through a hole at the center of the cell (2.5 cm from the exit aperture). Similarly, a small hole with windows at approximately 1 cm from the exit aperture provides optical access for the absorption probe laser beam. The copper block is partially extruded to create a long hole in the ablation laser propagation direction (so-called "Snorkel"), shown in Fig. \ref{cell}. This reduces the amount of ablation dust building up on the ablation window.

The buffer-gas cell is attached to the 4 K stage of a pulse tube refrigerator (PTR)(Sumitomo Heavy Industies SRP-062B) and held at $\sim$5 K. The inlet tube for helium buffer gas is thermally anchored to the 4 K stage of PTR to ensure pre-cooling to $\sim$5 K before entering the cell. Helium buffer gas is introduced from the inlet tube at the back of the cell and then passes through a diffuser at 3 mm from the gas inlet to further ensure good thermalization with the cell wall. Typical flow rate of helium buffer gas is $\sim$20 sccm. The helium buffer gas collides with and quickly thermalizes the ablated molecules, eventually leaving the cell through the exit aperture (5 mm diameter). Activated charcoal is attached on the interior surface of the 4 K radiation shields to cryo-pump buffer gas helium that comes out from the cell, maintaining high vacuum (typically $\sim$10$^{-5}$ torr while helium is flowing). The whole system is warmed up to the room temperature roughly every 5-7 days of active operation to release the gas trapped in the charcoal. The typical timescale of cooling down or warming up the system is $\sim$12 hours. The ability to easily attach and detach the target holder with screw allows quick powder target swapping. The powder target used in this work is commercially available. A single target with a size of $\sim$1 cm diameter yields $10^4$-$10^5$ shots. The signal size gradually decays when the ablation spot is unchanged, which is common in other species as well  (\cite{Iwata2017}).

All measurements in this work were performed on the molecules inside the buffer-gas cell, as opposed to the extracted molecular beam from the cell because of higher molecule density in the cell. A ring cavity dye laser (Coherent 899 dye laser, output power $\sim$500 mW, bandwidth $\sim$1 MHz) is used to excite the $\tilde{A}^2\Pi(0,0,0)-\tilde{X}^2\Sigma^+(0,0,0)$ transition of CaOH whose wavlength is $\sim$625 nm. The laser beam is then divided into three parts; a wavemeter, I$_2$ Doppler-free saturated absorption spectroscopy, and the main experiment (buffer-gas cell). The wavemeter (High finesse, WS6-200) has an absolute accuracy of 200 MHz. The absolute frequency is obtained by comparing the co-recorded I$_2$ spectra with the I$_2$ atlas  (\cite{I2_atlas_Baba}) and interpolating the lower and higher calibrated frequencies of each line. The absorption probe laser with a power of $\sim$100 $\mu$W and a diameter of $\sim$1 mm is sent perpendicular to the direction of the flow of molecular cloud toward the exit aperture to mitigate the systematic Doppler shift (indicated as red line in the Fig. \ref{cell}).

\section{Results}

\begin{figure}[ht]
\begin{center}
\includegraphics[width=0.48\textwidth]{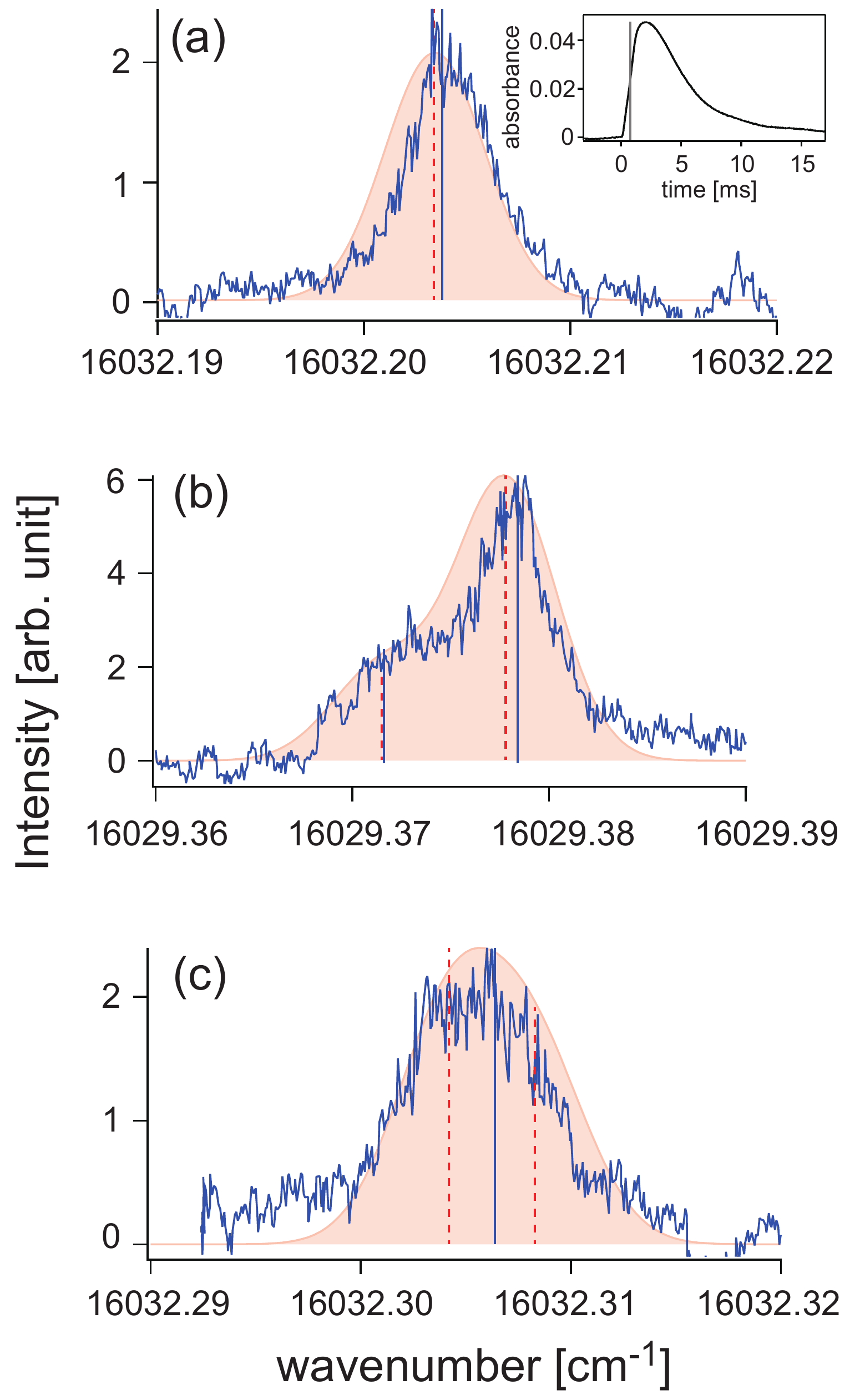}
\end{center}
\caption{Typical Doppler-broadened temporal absorption signal (inset) and spectra in three cases; (a) a single peak in an isolated region, (b) two partially-resolved
spin-rotation components ($Q_2(9/2)$ and $P_{21}(11/2)$), and (c) two unresolved spin-rotation components ($Q_{21}(7/2)$ and $R_2(5/2)$). The solid gray line in the inset indicate the integration time window. The red dashed (solid blue) lines indicate the fitted (observed) positions. The red area shows the simulated spectrum from the fit results at a translational temperature of 14 K.}
\label{spectra}
\end{figure}

A typical averaged absorption signal as a function of time from the ablation (t = 0) is shown in the inset of Fig. \ref{spectra}. After the rapid increase from t = 0 to 2 ms, the signal gradually decreases with long tail of \textgreater 10 ms. The time window for the time-integrated absorption signal is 500 ns, peaking at t = 0.8 ms. The narrow time window is chosen to make the signal less sensitive to the molecule flow dynamics which may exist in the cell. The typical in-cell CaOH density is 10$^8$-10$^9$ cm$^{-3}$ and shot-by-shot variation in signal size is around a factor of few. Relatively slow signal decay, with the time scale of $\sim$10 s (with 10 Hz repetition rate of ablation), is also present for each target spot.  The ablation laser position is therefore changed every $\sim$100 s (1000 shots).

Figure \ref{spectra} shows a typical Doppler-broadened absorption spectra. Transitions without spin-rotation splitting such as $R_{1}(J)$, $R_{21}(J)$, and transitions from the rotational ground state can be observed as isolated peaks (Fig. \ref{spectra} (a)). Other transitions split due to the spin-rotation interaction, which is partially unresolved in the present experiment due to Doppler width. As the splitting is comparable to the Doppler width, the resolved lines were also observed as shoulders (Fig. \ref{spectra} (b)) When the spin-rotation splitting is unresolved (Fig. \ref{spectra} (c)), the observed peak is fitted with a single Gaussian, and two spin-rotation components are assigned to the same observed line. The weaker transitions are ignored if the intensity ratio of unresolved transition pairs in the simulation at 10 K is less than 1/3. The two components obtained by the fitting are usually on both sides of the observed transition. The line residuals are smaller than the transition linewidth.
In total, 30 peaks for both $\Omega$ = 1/2 and $\Omega$ = 3/2 spin-orbit components are recorded. Two spin-orbit components are roughly spaced by 66 cm$^{-1}$. The linewidth of each line is about 170 MHz, which corresponds to the Doppler width at 14 K. Due to the shot-by-shot fluctuation in signal size, precise determination of rotational temperature is difficult, although the intense low-$J$ transitions suggest an approximate rotational temperature of 10 K.

The frequency is corrected with Doppler-free I$_2$ spectra, whose uncertainties are less than $\sim$10 MHz (\cite{I2_atlas_Baba}). The fitting errors of the CaOH spectra with a Gaussian function are also less than $\sim$10 MHz. Statistical uncertainty was estimated to be \textless 20 MHz by measuring each transition multiple times. One of the possible systematic errors is Doppler shift due to the flow of molecular clouds in the cell. In the previous simulation, the flow velocity toward the cell aperture is $\sim$10 m/s (\cite{Yuiki2021}). To get a conservative estimate, if we assume that the flow of $\sim$10 m/s is in the direction opposite to the laser beam, a shift of 16 MHz occurs. To investigate this systematic error, we checked the line positions using the probe absorption laser sent in the opposite direction. We found that the shift depends on the delay from ablation and is less than 15 MHz at 0.8 ms after the ablation. One of the possible explanations for this shift is the molecule flow dynamics in the cell. This systematic effect can be eliminated by using a technique of sending lasers from both directions, e.g. saturated absorption spectroscopy. More detail investigations on the cause of this error source will be included in a separate paper. Combined with the fitting error and uncertainty from multiple measurements, we estimated the total uncertainty of measurement to be \textless 30 MHz. This is an improvement in measurement uncertainty by more than a factor of three compared to \cite{Li1995, Dick2006}\footnote{The uncertainty in \cite{Dick2006} was estimated as "0.005 cm$^{-1}$ (150 MHz) for clean lines and 0.02 cm$^{-1}$ (600 MHz) for overlapped lines."}. 

The I$_2$ corrected transition wavenumbers and assignments are given in Table \ref{tab:freq} with fit results from PGOPHER program (\cite{Pgopher}). For the transition designation, a traditional $\Delta J_{F' F"} (J")$ notation was used (\cite{Herzberg1950}). A least-squares fit of the measured data to the Hund’s case (b) $\tilde{X}^2\Sigma^+$ – Hund’s case (a) $\tilde{A}^2\Pi$ Hamiltonian of Brown was performed in the PGOPHER program. Because low-$J$ transitions have a small dependency on higher order spectroscopic constants, centrifugal distortion constants $D"$ and $D'$, spin-rotation constant $\gamma "$, $\Lambda$-type doubling constant $q '$, and higher-order spin-orbit constant $A_D '$ were fixed to the previously obtained values (\cite{Dick2006}). (see Appendix \ref{fixed_para})  The standard deviation of the fitting residual is 0.0015 cm$^{-1}$ \footnote{The fitting residual is limited by roughly the spin-rotation splitting because it is partially unresolved as explained above. If we only select out spin-rotationally resolved lines, the standard deviation of the fit residual reduces to $\sim$0.0006 cm$^{-1}$ (18 MHz).}. As a result, the 30 observed transitions were assigned to the 40 rotational lines due to the unresolved spin-rotation splittings mentioned above. There are 27 transitions that have not been previously reported in literature and are assigned in this work.

The obtained spectroscopic parameters are shown in Table \ref{tab:constants}. The determined rotational constants, $B$, for both the ground $\tilde{X}^2\Sigma^+$ and excited $\tilde{A}^2\Pi$ states are in excellent agreement with previously determined values within 1 MHz. The difference in the band origin $T_{0}$ is $\sim$0.01 cm$^{-1}$ (300 MHz), which is comparable to the uncertainty of the previous works, but significant for the uncertainty in present measurement ($\sim$30 MHz). This indicates the improvement of the accuracy of the absolute transition frequency values. The difference in the spin-orbit constant $A'$ is larger than the present measurement uncertainty\footnote{Not to be confused with the fitting error.} whereas the $\Lambda$-type doubling constant $p'$ is comparable.

\begin{table*}[ht]
\caption{\label{tab:freq}
Observed lines, rotational states $J$, observed positions, calculated positions, residuals, previously measured positions (if present) (\cite{Dick2006}), and differences between this work and previous work in the $\tilde{A}^2\Pi(0,0,0)-\tilde{X}^2\Sigma^+(0,0,0)$ band of CaOH. The frequency unit is wave number (cm$^{-1}$). The $\Delta J_{F' F"}$ notation is used for specifying transitions. }
\begin{ruledtabular}
\begin{tabular}{llllll}
lines	&	$J"$	&	Observed	&	Obs-Calc	&	\cite{Dick2006}	&	Difference	 \\ \hline \\

$P_{1}$	&	3/2	&	15964.3929	&	0.0009	&	15964.38*	&	0.0129	 \\ 											
	&	5/2	&	15964.0968	&	0.0017	&	15964.091 	&	0.0058	 \\ 
	&	7/2	&	15963.8114	&	0.0028	&	15963.811 	&	0.0004	 \\ 
	
$P_{21}$ &	11/2	&	16029.3716	&	0.0001	&		&	\\ 	
	
$Q_{1}$	&	1/2	&	15965.0168	&	0.0002	&		&		 \\ 											
	&	3/2	&	15965.3445	&	0.0009	&		&		 \\ 
	&	5/2	&	15965.6821	&	0.0011	&		&		 \\ 
	&	7/2	&	15966.0308	&	0.0020	&		&		 \\ 
	&	9/2	&	15966.3899	&	0.0028	&		&		 \\ 
	&	13/2	&	15967.1383	&	0.0035	&		&		 \\ 
	
$Q_{2}$	&	3/2	&	16030.1986	&	-0.0007	&		&		 \\ 
	&	5/2	&	16029.9076	&	-0.0008	&	16029.933 	&	-0.0254	 \\ 
	&	7/2	&	16029.6344	&	-0.0001	&	16029.661 	&	-0.0266	 \\ 
	&	9/2	&	16029.3784	&	0.0006	&	16029.383 	&	-0.0046	 \\ 

$Q_{12}$	&	1/2	&	15964.3929	&	-0.0009	&		&		 \\ 											
	&	3/2	&	15964.0968	&	-0.0012	&		&		 \\ 
	&	5/2	&	15963.8114	&	-0.0013	&		&		 \\ 

$Q_{21}$	&	3/2	&	16031.5352	&	0.0010	&		&		 \\ 
	&	5/2	&	16031.9119	&	0.0012	&		&		 \\ 
	&	7/2	&	16032.3064	&	0.0022	&		&		 \\ 
	&	9/2	&	16032.7144	&	-0.0005	&		&		 \\ 
	&	11/2	&	16033.1421	&	-0.0004	&		&		 \\ 
	
$R_{1}$	&	1/2	&	15966.1017	&	-0.0005	&		&		 \\ 
	&	3/2	&	15967.1521	&	-0.0008	&		&		 \\ 
	&	5/2	&	15968.2129	&	-0.0011	&		&		 \\ 
	&	7/2	&	15969.2854	&	0.0000	&		&		 \\ 

$R_{2}$	&	1/2	&	16031.5352	&	-0.0008	&		&		 \\ 
	&	3/2	&	16031.9119	&	-0.0017	&	16031.931 	&	-0.0191	 \\ 
	&	5/2	&	16032.3064	&	-0.0019	&	16032.327 	&	-0.0206	 \\ 
	&	7/2	&	16032.7199	&	-0.0002	&	16032.734 	&	-0.0141	 \\ 
	&	9/2	&	16033.1486	&	-0.0003	&	16033.163 	&	-0.0144	 \\ 

$R_{12}$	&	3/2	&	15965.6821	&	-0.0018	&		&		 \\ 
	&	5/2	&	15966.0308	&	-0.0021	&		&		 \\ 
	&	7/2	&	15966.3899	&	-0.0024	&		&		 \\ 
	&	11/2	&	15967.1383	&	-0.0041	&		&		 \\

$R_{21}$	&	1/2	&	16032.2038	&	0.0004	&		&		 \\ 
	&	3/2	&	16033.2485	&	-0.0001	&		&		 \\ 
	&	5/2	&	16034.3106	&	-0.0002	&	16034.328 	&	-0.0174	 \\ 
	&	7/2	&	16035.3909	&	0.0009	&	16035.406 	&	-0.0151	 \\ 
	&	9/2	&	16036.4876	&	0.0013	&	16036.502 	&	-0.0144	 \\

\end{tabular}
\end{ruledtabular}
\begin{tablenotes}
\item{*} This value is from \cite{kozyryev_determination_2019-2}
\end{tablenotes}
\end{table*}

\begin{table*}[ht]
\caption{\label{tab:constants}
Electronic states, determined spectroscopic constants and its values, previously obtained values (\cite{Dick2006}), and difference between this work and previous work. The numbers in the parentheses indicate the fitting errors. The frequency unit is wave number (cm$^{-1}$). 
}
\begin{ruledtabular}
\begin{tabular}{lllll}
State	&	Constant	&	This work		&	\cite{Dick2006} 	&	Difference	 \\ 
\hline \\
$\tilde{X}^2\Sigma^+$	&	$B"$	&	0.334315(60)	&	0.334334107(32)	&	-1.9107 $\times$ 10$^{-5}$	 \\ 
$\tilde{A}^2\Pi$	&	$T_{0}$	&	15997.76559(44)	&	15997.77579(64)	&	-0.0102	 \\ 
	&	$B'$	&	0.341195(53)	&	0.3412272(15)	&	-3.22 $\times$ 10$^{-5}$	 \\ 
	&	$A'$	&	66.81849(53)	&	66.81480(89)	&	0.00369	 \\ 
	&	$p'$	&	-0.04469(23)	&	-0.0431(71)	&	-0.001626	 \\ 
                  
\end{tabular}
\end{ruledtabular}
\end{table*}

\section{Conclusion}

Using the BGC method, laboratory high-resolution spectra of the $\tilde{A}^2\Pi(0,0,0)-\tilde{X}^2\Sigma^+(0,0,0)$ band of CaOH were recorded. We assigned a total of 40 low-$J$ transitions, 27 of which do not exist in current previous literature. Rotational constants were determined for
both $\tilde{A}^2\Pi(0,0,0)$ and $\tilde{X}^2\Sigma^+(0,0,0)$ states and are consistent with previous measurements. In contrast, the experimental uncertainty for the absolute frequencies of low-$J$ transitions was improved by more than a factor of three. The transition frequencies along with spectroscopic constants obtained in this work are expected to contribute to the MARVEL analysis as additional data set, aiding in the identification of interstellar, circumstellar, and atmospheric CaOH. The ability to cool the molecular species down to Kelvin regime with BGC method would also be useful for probing low-$J$ transitions of other astrophysical molecules, especially in an interstellar environment where the temperature can be as cold as a few Kelvins. It can also be combined with microwave spectroscopy (\cite{Porterfield2019}) to probe the low temperature chemical reactions, opening up a new path to study chemistry in the intersteller medium.

\begin{acknowledgments}

We acknowledge many helpful discussions and kind cooperation from Yosuke Takasu, Satoshi Uetake, Nick Hutzler, Christian Hallas, Louis Baum, Ben Augenbraun, Yoshiro Takahashi, and John Doyle. Y. T. was supported by the Masason Foundation. This work was supported by the Masason Foundation and JSPS KAKENHI Grant No. 18H01229, 22H01249.

\end{acknowledgments}

\appendix

\section{\label{fixed_para}Fixed parameters in the fit}

\begin{table}[ht]
\caption{\label{tab:fixed_constants}
Electronic states, fixed spectroscopic constants in the fit and its values. The constants are fixed to the obtained values in \cite{Dick2006}. The frequency unit is wave number (cm$^{-1}$). 
}
\begin{ruledtabular}
\begin{tabular}{lll}
State	&	Constant	&	Fixed value	 \\ 
\hline \\
$\tilde{X}^2\Sigma^+$	&$D"$	&	3.86 $\times$ 10$^{-7}$	 \\ 
	&	$\gamma "$	&	0.00115957	 \\ 
$\tilde{A}^2\Pi$	&	$D'$	&	3.896 $\times$ 10$^{-7}$	 \\ 
	&	$q'$	&	-3.447 $\times$ 10$^{-6}$	 \\ 
	&	$A_D'$	&	-1.741 $\times$ 10$^{-4}$	 \\ 
                  
\end{tabular}
\end{ruledtabular}
\end{table}

\bibliography{biblio,ref2}

\begin{thebibliography}{}
\expandafter\ifx\csname natexlab\endcsname\relax\def\natexlab#1{#1}\fi
\providecommand{\url}[1]{\href{#1}{#1}}
\providecommand{\dodoi}[1]{doi:~\href{http://doi.org/#1}{\nolinkurl{#1}}}
\providecommand{\doeprint}[1]{\href{http://ascl.net/#1}{\nolinkurl{http://ascl.net/#1}}}
\providecommand{\doarXiv}[1]{\href{https://arxiv.org/abs/#1}{\nolinkurl{https://arxiv.org/abs/#1}}}

\bibitem[{Baum {et~al.}(2020{\natexlab{a}})Baum, Vilas, Hallas, Augenbraun,
  Rava, Mitra, \& Doyle}]{baum_establishing_2020}
Baum, L., Vilas, N.~B., Hallas, C., {et~al.} 2020{\natexlab{a}},
  arXiv:2006.01769.
\newblock \url{http://arxiv.org/abs/2006.01769}

\bibitem[{Baum {et~al.}(2020{\natexlab{b}})Baum, Vilas, Hallas, Augenbraun,
  Raval, Mitra, \& Doyle}]{baum_1d_2020-1}
---. 2020{\natexlab{b}}, Physical Review Letters, 124, 133201,
  \dodoi{10.1103/PhysRevLett.124.133201}

\bibitem[{Bernath \& Kinsey-Nielsen(1984)}]{Bernath1984}
Bernath, P., \& Kinsey-Nielsen, S. 1984, Chemical Physics Letters, 105, 663,
  \dodoi{https://doi.org/10.1016/0009-2614(84)85678-X}

\bibitem[{Bernath(2009)}]{Bernath2009}
Bernath, P.~F. 2009, International Reviews in Physical Chemistry, 28, 681,
  \dodoi{10.1080/01442350903292442}

\bibitem[{Bernath \& Brazier(1985)}]{Bernath1985}
Bernath, P.~F., \& Brazier, C.~R. 1985, Astrophys. J., 288, 373,
  \dodoi{10.1086/162800}

\bibitem[{Changala {et~al.}(2019)Changala, Weichman, Lee, Fermann, \&
  Ye}]{Changala2019}
Changala, P.~B., Weichman, M.~L., Lee, K.~F., Fermann, M.~E., \& Ye, J. 2019,
  Science, 363, 49, \dodoi{10.1126/science.aav2616}

\bibitem[{Coxon {et~al.}(1994)Coxon, Li, \& Presunka}]{Coxon1994}
Coxon, J., Li, M., \& Presunka, P. 1994, J. Mol. Spectrosc., 164, 118,
  \dodoi{10.1006/jmsp.1994.1060}

\bibitem[{Coxon {et~al.}(1991)Coxon, Li, \& Presunka}]{Coxon1991}
Coxon, J.~A., Li, M., \& Presunka, P.~I. 1991, Journal of Molecular
  Spectroscopy, 150, 33, \dodoi{https://doi.org/10.1016/0022-2852(91)90191-C}

\bibitem[{Coxon {et~al.}(1992)Coxon, Li, \& Presunka}]{Coxon1992}
---. 1992, Molecular Physics, 76, 1463, \dodoi{10.1080/00268979200102231}

\bibitem[{Császár {et~al.}(2007)Császár, Czakó, Furtenbacher, \&
  Mátyus}]{CSASZAR2007}
Császár, A.~G., Czakó, G., Furtenbacher, T., \& Mátyus, E. 2007, in Annual
  Reports in Computational Chemistry, Vol.~3, Chapter 9 An Active Database
  Approach to Complete Rotational–Vibrational Spectra of Small Molecules, ed.
  D.~Spellmeyer \& R.~Wheeler (Elsevier), 155--176,
  \dodoi{https://doi.org/10.1016/S1574-1400(07)03009-5}

\bibitem[{Dick {et~al.}(2006)Dick, Sheridan, Wang, Yu, \& Bernath}]{Dick2006}
Dick, M., Sheridan, P., Wang, J.-G., Yu, S., \& Bernath, P. 2006, Journal of
  Molecular Spectroscopy, 240, 238,
  \dodoi{https://doi.org/10.1016/j.jms.2006.10.009}

\bibitem[{Furtenbacher \& Császár(2012)}]{Furtenbacher2012}
Furtenbacher, T., \& Császár, A.~G. 2012, Journal of Molecular Structure,
  1009, 123, \dodoi{https://doi.org/10.1016/j.molstruc.2011.10.057}

\bibitem[{Furtenbacher {et~al.}(2007)Furtenbacher, Császár, \&
  Tennyson}]{Furtenbacher2007}
Furtenbacher, T., Császár, A.~G., \& Tennyson, J. 2007, Journal of Molecular
  Spectroscopy, 245, 115, \dodoi{https://doi.org/10.1016/j.jms.2007.07.005}

\bibitem[{Herzberg(1950)}]{Herzberg1950}
Herzberg, G. 1950, Molecular Spectra and Molecular Structure, 2nd Ed. (Krieger
  Publishing Comapany)

\bibitem[{Hilborn {et~al.}(1983)Hilborn, Qingshi, \& Harris}]{Hilborn1983}
Hilborn, R.~C., Qingshi, Z., \& Harris, D.~O. 1983, Journal of Molecular
  Spectroscopy, 97, 73, \dodoi{https://doi.org/10.1016/0022-2852(83)90338-7}

\bibitem[{Hutzler {et~al.}(2012)Hutzler, Lu, \& Doyle}]{Hutzler2012}
Hutzler, N.~R., Lu, H.-I., \& Doyle, J.~M. 2012, Chem. Rev., 112, 4803,
  \dodoi{10.1021/cr200362u}

\bibitem[{Iwata {et~al.}(2017)Iwata, McNally, \& Zelevinsky}]{Iwata2017}
Iwata, G.~Z., McNally, R.~L., \& Zelevinsky, T. 2017, Phys. Rev. A, 96, 022509,
  \dodoi{10.1103/PhysRevA.96.022509}

\bibitem[{Jarman \& Bernath(1992)}]{Jarman1992}
Jarman, C.~N., \& Bernath, P.~F. 1992, J. Chem. Phys., 97, 1711,
  \dodoi{10.1063/1.463158}

\bibitem[{Kato {et~al.}(2000)Kato, Kasahara, Misono, \& Baba}]{I2_atlas_Baba}
Kato, H., Kasahara, S., Misono, M., \& Baba, M. 2000, JSPS

\bibitem[{Kozyryev {et~al.}(2019)Kozyryev, Steimle, Yu, Nguyen, \&
  Doyle}]{kozyryev_determination_2019-2}
Kozyryev, I., Steimle, T.~C., Yu, P., Nguyen, D.-T., \& Doyle, J.~M. 2019, New
  Journal of Physics, 21, 052002, \dodoi{10.1088/1367-2630/ab19d7}

\bibitem[{Li \& Coxon(1992)}]{li_laser_1992}
Li, M., \& Coxon, J.~A. 1992, The Journal of Chemical Physics, 97, 8961,
  \dodoi{10.1063/1.463322}

\bibitem[{Li \& Coxon(1995)}]{Li1995}
---. 1995, J. Chem. Phys., 102, 2663, \dodoi{10.1063/1.468643}

\bibitem[{Maxwell {et~al.}(2005)Maxwell, Brahms, DeCarvalho, Glenn, Helton,
  Nguyen, Patterson, Petricka, DeMille, \& Doyle}]{Maxwell2005}
Maxwell, S.~E., Brahms, N., DeCarvalho, R., {et~al.} 2005, Phys. Rev. Lett.,
  95, 173201, \dodoi{10.1103/PhysRevLett.95.173201}

\bibitem[{McCarron(2018)}]{McCarron2018}
McCarron, D. 2018, J. Phys. B At. Mol. Opt. Phys., 51, 212001,
  \dodoi{10.1088/1361-6455/aadfba}

\bibitem[{Messer \& {De Lucia}(1984)}]{Messer1984}
Messer, J.~K., \& {De Lucia}, F.~C. 1984, Phys. Rev. Lett., 53, 2555,
  \dodoi{10.1103/PhysRevLett.53.2555}

\bibitem[{Miyamoto {et~al.}(2022)Miyamoto, Tobaru, Takahashi, Hiramoto,
  Iwakuni, Kuma, Enomoto, \& Baba}]{Miyamoto2022arxiv}
Miyamoto, Y., Tobaru, R., Takahashi, Y., {et~al.} 2022, arXiv,
  \dodoi{10.48550/ARXIV.2206.07927}

\bibitem[{Owens {et~al.}(2021)Owens, Clark, Mitrushchenkov, Yurchenko, \&
  Tennyson}]{Owens2021}
Owens, A., Clark, V. H.~J., Mitrushchenkov, A., Yurchenko, S.~N., \& Tennyson,
  J. 2021, The Journal of Chemical Physics, 154, 234302,
  \dodoi{10.1063/5.0052958}

\bibitem[{Porterfield {et~al.}(2019)Porterfield, Satterthwaite, Eibenberger,
  Patterson, \& McCarthy}]{Porterfield2019}
Porterfield, J.~P., Satterthwaite, L., Eibenberger, S., Patterson, D., \&
  McCarthy, M.~C. 2019, Review of Scientific Instruments, 90, 053104,
  \dodoi{10.1063/1.5091773}

\bibitem[{{Rajpurohit, A. S.} {et~al.}(2016){Rajpurohit, A. S.}, {Reyl\'e, C.},
  {Allard, F.}, {Homeier, D.}, {Bayo, A.}, {Mousis, O.}, {Rajpurohit, S.}, \&
  {Fern\'andez-Trincado, J. G.}}]{Rajpurohit2016}
{Rajpurohit, A. S.}, {Reyl\'e, C.}, {Allard, F.}, {et~al.} 2016, A\&A, 596,
  A33, \dodoi{10.1051/0004-6361/201526776}

\bibitem[{{Rajpurohit, A. S.} {et~al.}(2013){Rajpurohit, A. S.}, {Reyl\'e, C.},
  {Allard, F.}, {Homeier, D.}, {Schultheis, M.}, {Bessell, M. S.}, \& {Robin,
  A. C.}}]{Rajpurohit2013}
---. 2013, A\&A, 556, A15, \dodoi{10.1051/0004-6361/201321346}

\bibitem[{Santamaria {et~al.}(2016)Santamaria, Sarno, Natale, Rosa, Inguscio,
  Mosca, Ricciardi, Calonico, Levi, \& Maddaloni}]{Santamaria2016}
Santamaria, L., Sarno, V.~D., Natale, P.~D., {et~al.} 2016, Phys. Chem. Chem.
  Phys., 18, 16715, \dodoi{10.1039/C6CP02163H}

\bibitem[{Scurlock {et~al.}(1993)Scurlock, Fletcher, \& Steimle}]{Scurlock1993}
Scurlock, C., Fletcher, D., \& Steimle, T. 1993, Journal of Molecular
  Spectroscopy, 159, 350, \dodoi{https://doi.org/10.1006/jmsp.1993.1133}

\bibitem[{Spaun {et~al.}(2016)Spaun, Changala, Patterson, Bjork, Heckl, Doyle,
  \& Ye}]{Spaun2016}
Spaun, B., Changala, P.~B., Patterson, D., {et~al.} 2016, Nature, 533, 517,
  \dodoi{10.1038/nature17440}

\bibitem[{Steimle {et~al.}(1992)Steimle, Fletcher, Jung, \&
  Scurlock}]{steimle_supersonic_1992}
Steimle, T.~C., Fletcher, D.~a., Jung, K.~Y., \& Scurlock, C.~T. 1992, The
  Journal of Chemical Physics, 96, 2556, \dodoi{10.1063/1.462007}

\bibitem[{Takahashi {et~al.}(2021)Takahashi, Shlivko, Woolls, \&
  Hutzler}]{Yuiki2021}
Takahashi, Y., Shlivko, D., Woolls, G., \& Hutzler, N.~R. 2021, Phys. Rev.
  Research, 3, 023018, \dodoi{10.1103/PhysRevResearch.3.023018}

\bibitem[{Tarbutt(2018)}]{Tarbutt2019}
Tarbutt, M.~R. 2018, Contemp. Phys., 59, 356,
  \dodoi{10.1080/00107514.2018.1576338}

\bibitem[{Tennyson \& Yurchenko(2017)}]{Tennyson2017}
Tennyson, J., \& Yurchenko, S.~N. 2017, Molecular Astrophysics, 8, 1,
  \dodoi{https://doi.org/10.1016/j.molap.2017.05.002}

\bibitem[{Tóbiás {et~al.}(2019)Tóbiás, Furtenbacher, Tennyson, \&
  Császár}]{Tobias2019}
Tóbiás, R., Furtenbacher, T., Tennyson, J., \& Császár, A.~G. 2019, Phys.
  Chem. Chem. Phys., 21, 3473, \dodoi{10.1039/C8CP05169K}

\bibitem[{{V. Andreev} {et~al.}(2018){V. Andreev}, Ang, DeMille, Doyle,
  Gabrielse, Haefner, Hutzler, Lasner, Meisenhelder, O'Leary, Panda, West,
  West, \& Wu}]{ACME2018}
{V. Andreev}, Ang, D.~G., DeMille, D., {et~al.} 2018, Nature, 562, 355,
  \dodoi{10.1038/s41586-018-0599-8}

\bibitem[{Vilas {et~al.}(2022)Vilas, Hallas, Anderegg, Robichaud, Winnicki,
  Mitra, \& Doyle}]{Vilas2022}
Vilas, N.~B., Hallas, C., Anderegg, L., {et~al.} 2022, Nature, 606, 70,
  \dodoi{10.1038/s41586-022-04620-5}

\bibitem[{Wang {et~al.}(2020)Wang, Owens, Tennyson, \& Yurchenko}]{Wang2020}
Wang, Y., Owens, A., Tennyson, J., \& Yurchenko, S.~N. 2020, The Astrophysical
  Journal Supplement Series, 248, 9, \dodoi{10.3847/1538-4365/ab85cb}

\bibitem[{{Western}(2017)}]{Pgopher}
{Western}, C.~M. 2017, \jqsrt, 186, 221, \dodoi{10.1016/j.jqsrt.2016.04.010}

\bibitem[{{Ziurys} {et~al.}(1992){Ziurys}, {Barclay}, \&
  {Anderson}}]{Ziurys1992}
{Ziurys}, L.~M., {Barclay}, W.~L., J., \& {Anderson}, M.~A. 1992, \apjl, 384,
  L63, \dodoi{10.1086/186262}

\bibitem[{{Ziurys} {et~al.}(1996){Ziurys}, {Fletcher}, {Anderson}, \&
  {Barclay}}]{Ziurys1996}
{Ziurys}, L.~M., {Fletcher}, D.~A., {Anderson}, M.~A., \& {Barclay}, W.~L., J.
  1996, \apjs, 102, 425, \dodoi{10.1086/192265}

\end{thebibliography}

\end{document}